\documentstyle[12pt]{article}
\begin{document}

\def\beq{\begin{equation}}
\def\eeq{\end{equation}}
\def\bce{\begin{center}}
\def\ece{\end{center}}
\def\bea{\begin{eqnarray}}
\def\eea{\end{eqnarray}}
\def\ben{\begin{enumerate}}
\def\een{\end{enumerate}}
\def\ul{\underline}
\def\ni{\noindent}
\def\nn{\nonumber}
\def\bs{\bigskip}
\def\ms{\medskip}
\def\wt{\widetilde}
\def\wh{\widehat}
\def\brr{\begin{array}}
\def\err{\end{array}}
\def\dsp{\displaystyle}

\hfill IEEC 97-71

\hfill hep-th/9707257

\hfill July 1997

\vspace*{15mm}

\begin{center}

{\LARGE \bf Multidimensional extension of the generalized
Chowla--Selberg formula}

\vspace{12mm}

\medskip

{\sc E. Elizalde}\footnote{E-mail address:
eli@zeta.ecm.ub.es, elizalde@io.ieec.fcr.es}\\
Consejo Superior de Investigaciones Cient\'{\i}ficas (CSIC),\\
Institut d'Estudis Espacials de Catalunya (IEEC), \\
Edifici Nexus 104, Gran Capit\`a 2-4, 08034 Barcelona, Spain\\ and \\
Departament ECM and IFAE, Facultat de F\'{\i}sica, \\
Universitat de Barcelona, Diagonal 647,
08028 Barcelona, Spain \\

\vspace{20mm}

{\bf Abstract}

\end{center}

After recalling the precise existence conditions
of the zeta function
of a pseudodifferential operator, and the concept of reflection
formula,
an exponentially convergent expression for the analytic
continuation of a multidimensional
inhomogeneous Epstein-type  zeta function of the general form
\[
\zeta_{A,\vec{b},q} (s) = \sum_{\vec{n}\in \mbox{\bf Z}^p} (\vec{n}^T A \vec{n}
+\vec{b}^T \vec{n}+q)^{-s}, \]
with $A$ the $p\times p$ matrix of a quadratic form, $\vec{b}$ a $p$
vector and $q$ a constant,
is obtained. It is valid on the whole complex $s$-plane,
is exponentially convergent and provides
the residua at the poles explicitly. It reduces to the famous
formula of Chowla and Selberg in the particular case $p=2$,
$\vec{b}= \vec{0}$, $q=0$. Some variations of the formula and 
physical applications are considered.

\vfill


%\noindent {\it Math. Sub. Class.:} 11M41, 11M35, 30B50, 30B40

%\noindent {\it Keywords:} Zeta function, analytic continuation,
%Chowla--Selberg formula, determinant, effective action

\newpage

\noindent{\large \bf 1. Introduction:
Existence of the zeta function
of a pseudodifferential operator ($\Psi$DO)}

A {\it pseudodifferential operator} $A$ of order
$m$ on a manifold $M_n$ is defined through its symbol
$a(x,\xi)$, which is a  function  belonging to the  space
$S^m(\mbox{\bf R}^n\times \mbox{\bf R}^n)$
of $\mbox{\bf C}^\infty$ functions
such that for any pair of multi-indexs $\alpha, \beta$ there
exists a constant $C_{\alpha,\beta}$ so that
\begin{equation}
\left| \partial^\alpha_\xi \partial^\beta_x a(x,\xi) \right| \leq
 C_{\alpha,\beta} (1+|\xi|)^{m-|\alpha|}.
\end{equation}
The definition of $A$
is given, in the distribution sense, by
\begin{equation}
Af(x) = (2\pi)^{-n} \int e^{i<x,\xi>} a(x,\xi) \hat{f}(\xi) \, d\xi,
\end{equation}
where $f$ is a smooth function, $f \in
 {\cal S}$  [remember that ${\cal S} = \left\{ f \in  C^\infty
(\mbox{\bf R}^n);  \mbox{sup}_x |x^\beta \partial^\alpha f(x) |
< \infty, \right.$  $\left. \forall \alpha, \beta \in \mbox{\bf
R}^n\right\}$], $ {\cal S}'$ being the space of tempered distributions
and $\hat{f}$ the Fourier transform of $f$.
 When $a(x,\xi)$ is a polynomial in $\xi$
 one gets a differential operator.
In general,  the order $m$ can be complex.
The {\it symbol} of a $\Psi$DO  has the form
\begin{equation}
a (x,\xi) =a _m(x,\xi) +a _{m-1}(x,\xi) + \cdots
+a _{m-j}(x,\xi) + \cdots, \label{spsd}
\end{equation}
being $a _k(x,\xi) = b_k(x) \, \xi^k$.

  Pseudodifferential
operators are useful tools, both in mathematics and in physics.
 They were crucial for the proof of the
uniqueness of the Cauchy problem \cite{cald57}
and also for the proof of the Atiyah-Singer index formula \cite{as63}.
In quantum field theory they appear in any
 analytical continuation process (as
complex powers of differential
operators, like the Laplacian) \cite{seel1}. And
they constitute nowadays the basic starting point
of any rigorous formulation of quantum
field theory through microlocalization,
a concept that is considered to be
 the most important step towards the understanding
of linear partial differential equations since the invention
of distributions  \cite{psdo1}.

For $A$ a positive-definite elliptic $\Psi$DO of
positive order $m \in \mbox{\bf R}$, acting on
the space of smooth sections of an
$n$-dimensional vector bundle $E$ over a
 closed, $n$-dimensional
  manifold $M$, the {\it zeta function} is defined as
\begin{equation}
\zeta_A (s) = \mbox{tr}\  A^{-s} = \sum_j
 \lambda_j^{-s}, \qquad \mbox{Re}\ s>\frac{n}{m} \equiv s_0.
\end{equation}
The quotient $s_0=$ dim$\,M/$ord$\,A$ is called the {\it abscissa of
convergence} of $\zeta_A(s)$, which is proven to have a meromorphic
continuation to  the whole complex plane
$\mbox{\bf C}$ (regular at $s_0$),
provided that the principal symbol of $A$ (that is
$a_m(x,\xi)$) admits a {\it spectral cut}: $
L_\theta = \left\{  \lambda \in \mbox{\bf C};
 \mbox{Arg}\, \lambda =\theta,
\theta_1 < \theta < \theta_2\right\}$,   $\mbox{Spec}\, A
\cap L_\theta = \emptyset$
(the Agmon-Nierenberg condition). Strictly
 speaking, the definition of $\zeta_A (s)$ depends on the
position
of the cut $L_\theta$, not so that of the
 determinant \cite{rs1} $\det_\zeta A = \exp
 [-\zeta_A'(0)]$, which only depends on the
homotopy class of the cut. The precise structure
of the analytical continuation of the zeta function is known in general
\cite{gilk1}. The only singularities it can have are
{\it simple poles} at
\begin{equation}
s_k = (n-k)/m,  \qquad  k=0,1,2,\ldots,n-1,n+1, \dots.
\end{equation}
The applications of the zeta-function definition
of a determinant in physics are very important \cite{sz1,wit1}.
 A zeta function with the same
meromorphic structure in the complex $s$-plane
and extending the ordinary definition
to operators of complex
order $m\in \mbox{\bf C} \backslash  \mbox{\bf Z}$
(it is clear that operators of complex order do not
admit spectral cuts), has been obtained in Ref. \cite{kv1}.
 The  construction  starts there from
 the definition of a trace, obtained as the integral
over the manifold of the trace density of the difference
between the Schwartz kernel of $A$ and the Fourier
transformed of a number of first homogeneous terms (in
$\xi$) of the usual decomposition of the symbol (\ref{spsd}) of $A$.

\bigskip

\noindent{\large \bf 2. Exponentially convergent analytic
continuation (with explicit poles and residua) of our zeta function}

A fundamental property shared by zeta functions of any nature is the
existence
of a reflection formula. For a generic zeta function, $Z(s)$, it has
the form:
\beq
Z(\omega -s)= F(\omega, s) Z(s),
\eeq
 and allows for its analytical continuation in a very
easy way ---what is the whole story of the zeta function regularization
 procedure (or at least the main part of it). But the
analytically continued expression thus obtained
 is just another series, which has again
a slow convergence behavior, of power series type (actually the same
that the original series had,
in its own domain of validity). Some years ago, the notorious
mathematicians S. Chowla and A. Selberg
found a formula, for the case $p=2$, $\vec{b}=0$ and
$q=0$ above \cite{cs},
that yields {\it exponentially quick convergence, and not only in the reflected
domain}. They were extremely proud of that formula ---as one can
appreciate by
reading the original paper, where actually no hint about its
 derivation was given.
In Ref. \cite{eli2a} we generalized this expression to inhomogeneous
zeta functions, but staying always in {\it two} dimensions ($p=2$),
for this was commonly believed to be an
unsurmountable restriction of the original formula (see, for instance,
Ref. \cite{dic}). The generalization we carried out in  \cite{eli2a}
 was already non trivial and a very detailed account of this step is
given in \cite{eli2} (where a missprint of the original derivation is
corrected). Finally, we have realized
that an extension to an {\it arbitrary} number of dimensions is
actually
possible. In fact, here we shall obtain formulas for arbitrary $p$
and arbitrary values of $\vec{b}$ and $q$.

The starting point will be {\it Poisson's resummation formula} in $p$
 dimensions, which arises from the distribution identity
\beq
\sum_{\vec{n} \in \mbox{\bf Z}^p} \delta (\vec{x}-\vec{n}) =
\sum_{\vec{m} \in \mbox{\bf Z}^p} e^{i 2\pi \vec{m} \cdot \vec{x}}.
\label{prs1}
\eeq
(We shall indistinctly write $ \vec{m} \cdot \vec{x} \equiv
 \vec{m}^T \vec{x}$ in what follows.) Applying this identity to the
function
\beq
f(\vec{x}) = \exp \left( -\frac{1}{2}\vec{x}^T A  \vec{x} +
 \vec{b}^T \vec{x} \right),
\eeq
with $A$ an invertible $p\times p$ matrix, and integrating then over
$\vec{x} \in \mbox{\bf R}^p$, one obtains
\beq
\sum_{\vec{n} \in \mbox{\bf Z}^p} \exp \left( -\frac{1}{2}\vec{n}^T A  \vec{n} +
 \vec{b}^T \vec{n} \right) = \frac{(2 \pi)^{p/2}}{\sqrt{\det A}}
\sum_{\vec{m} \in \mbox{\bf Z}^p}
 \exp \left[ \frac{1}{2}\left(\vec{b} + 2\pi i \vec{m} \right)^T A^{-1}
\left(\vec{b} + 2\pi i \vec{m} \right)\right].\label{prs2}
\eeq
We are going to consider the following zeta function
\beq
\zeta_{A,\vec{c},q} (s) = \sum_{\vec{n} \in \mbox{\bf Z}^p}' \left[
\frac{1}{2}\left( \vec{n}+\vec{c}\right)^T A
\left( \vec{n}+\vec{c}\right)+ q\right]^{-s} \equiv
 \sum_{\vec{n} \in \mbox{\bf Z}^p}'
\left[ Q\left( \vec{n}+\vec{c}\right)+ q\right]^{-s} , \ \ \
 \mbox{Re}\ (s) > \frac{p}{2}. \label{zf1}
\eeq
The aim is to obtain a formula that gives (the analytical continuation
of) this multidimensional
zeta function in terms of an exponentially convergent multiseries
and which is valid in
the whole complex plane, exhibiting the singularities (simple poles) of
the meromorphic continuation ---with the corresponding residua---
explicitly. The prime in the summatories of Eq. (\ref{zf1}) means
that the point $ \vec{n}=\vec{0}$ is to be excluded from the sum.
Such restriction is irrelevant as long as $q\neq 0$ (the contribution of
this single point being immediately obtainable), but
to define the zeta function in this way is essential
in order to be able to reach the limit $q \rightarrow 0$ (as we
shall do later). The only condition on the matrix $A$ is that
it corresponds to a (non negative) quadratic form, that we call $Q$.
The vector $\vec{c}$ is arbitrary, while $q$ will (for the
moment) be a positive constant.
\ms

\ni{\sf 2.1. The main expression (inhomogeneous case).}
Use of the Poisson resummation formula (\ref{prs2})
yields, after some work, the following expression:
\bea
\zeta_{A,\vec{c},q} (s) &=& \frac{(2\pi )^{p/2} q^{p/2 -s}}{
\sqrt{\det A}} \, \frac{\Gamma(s-p/2)}{\Gamma (s)} +
 \frac{2^{s/2+p/4+2}\pi^s q^{-s/2 +p/4}}{\sqrt{\det A} \
\Gamma (s)} \nn \\ && \times \sum_{\vec{m} \in
\mbox{\bf Z}^p_{1/2}}' \cos
(2\pi
 \vec{m}\cdot \vec{c}) \left( \vec{m}^T A^{-1} \vec{m}
\right)^{s/2-p/4} \, K_{p/2-s} \left( 2\pi \sqrt{2q \,
 \vec{m}^T A^{-1} \vec{m}}\right), \label{qpd1}
\eea
where $K_\nu$ is the modified Bessel function of the second kind and
the subindex 1/2 in
$\mbox{\bf Z}^p_{1/2}$ means that only half of the vectors
$\vec{m} \in \mbox{\bf Z}^p$ intervene in the sum. That is, if we take
an $\vec{m} \in \mbox{\bf Z}^p$
we must then exclude $-\vec{m}$ (as simple criterion one can,
for instance, select those vectors in {\bf Z}$^p \backslash \{ \vec{0}
\}$ whose first non-zero component is positive). Eq. (\ref{qpd1})
fulfills {\it all} the requirements demanded before. It
is notorious to observe how the only pole of this inhomogeneous Epstein
zeta function appears explicitly at $s=p/2$, where it belongs. Its
residue is given by the formula:
\beq
\mbox{Res}_{s=p/2}  \zeta_{A,\vec{c},q} (s) =
\frac{(2\pi )^{p/2}}{\sqrt{\det A} \ \Gamma(p/2)}.
\eeq
With a bit of
care, it is relatively simple to obtain the limit of expression
(\ref{qpd1}) as $q\rightarrow 0$.

However, instead of proceeding in this way (what we shall do later),
it is adviceable to construct first a direct recurrent formula for the
case $q=0$. This is certainly
the natural option in such case, where no cut-off $q$ exists
to safeguard the $t$-integration (there is {\it no} way to use the
Poisson formula
on all $p$ indices of $\vec{n}$ at once). However, we can still deal
with
this case by using the Poisson resummation formula on {\it some} of the $p$
indices $\vec{n}$ only, say on just one of them, $n_1$. 
Poisson's formula on one index
reduces to the celebrated Jacobi identity for the $\theta_3$ function
\begin{equation}
\theta_3 \left(z\left|\tau\right.\right) = 1 + 2
 \sum_{n=1}^{\infty} q^{n^2} \cos (2nz), \ \ \ \ \  q=e^{\pi i \tau}, \
\ |q| <1, \ \ \tau \in \mbox{\bf C}, \label{tf3d}
\end{equation}
the identity being:
\begin{equation}
\theta_3 \left(z\left|\tau\right.\right) =\frac{1}{\sqrt{-i\tau}}
 e^{z^2/( \pi  i \tau)}\theta_3
\left(\frac{z}{\tau}\left|\frac{-1}{\tau}\right.\right), \label{tfi1}
\end{equation}
or, in other words
\begin{equation}
\sum_{n=-\infty}^{\infty} e^{n^2\pi i \tau +2niz} =
 \frac{1}{\sqrt{-i\tau}} \,\sum_{n=-\infty}^{\infty}
e^{(z-n\pi)^2/( \pi i \tau)}\label{tfi2}
\end{equation}
(for a classical reference see, e.g. Ref.
\cite{ww}).
 Here $z$ and
$\tau$ are arbitrary complex,  $z,\tau \in$ {\bf C},  with the only
restriction that Im $\tau > 0$ (in order that $|q| <1$).
For the applications, it turns out to be better to recast
the Jacobi identity  as
follows (with $\pi i \tau \rightarrow  -t$ and $z \rightarrow \pi z$):
\begin{equation}
\sum_{n=-\infty}^{\infty} e^{-n^2t +2\pi i nz} =
 \sqrt{\frac{\pi}{t}} \,\sum_{n=-\infty}^{\infty}
e^{-\pi^2(n-z)^2/t },\label{tfi3}
\end{equation}
equivalently
 \begin{equation}
\sum_{n=-\infty}^{\infty} e^{-(n+z)^2t} =
 \sqrt{\frac{\pi}{t}} \left[ 1+ \sum_{n=1}^{\infty}
e^{-\pi^2n^2/t } \cos (2\pi n z) \right],\label{tfi4}
\end{equation}
with $z,t \in$ {\bf C}, Re $t>0$.
\ms

\ni{\sf 2.2. Recurrent expression (and the homogeneous case).}
Using this last formula on the first component, $n_1$, of the summation
vector
$\vec{n}$, we obtain (for the sake of simplicity of the final
expressions,
we shall just consider now the case $\vec{c} =\vec{0}$, but the result can
be generalized to $\vec{c} \neq \vec{0}$ quite easily):
 \begin{eqnarray}
&& \hspace{-5mm}  \zeta_{A,\vec{0},q} (s) = 2 \sum_{n_1=1}^\infty
\left( an_1^2+q\right)^{-s} + \frac{1}{\Gamma (s)}
\sum_{\vec{n}_2 \in \mbox{\bf Z}^{p-1}}' \left[  \sqrt{\frac{\pi}{a}}\,
\Gamma (s-1/2) \left( \vec{n}_2^T \Delta_{p-1}  \vec{n}_2
+q \right)^{1/2 -s} \right. \\ &&  \hspace{-10mm} \left.
+\frac{4 \pi^s}{a^{s/2+1/4}}
\sum_{n_1=1}^\infty
 \cos \left( \frac{\pi n_1}{a}  \vec{b}^T  \vec{n}_2 \right)
n_1^{s-1/2}  \left( \vec{n}_2^T \Delta_{p-1}  \vec{n}_2
+q \right)^{1/4 -s/2}
 K_{s-1/2} \left( \frac{2\pi n_1}{\sqrt{a}}
\sqrt{\vec{n}_2^T \Delta_{p-1}  \vec{n}_2 +q} \right) \right], \nn
\end{eqnarray}
which can be written as
 \begin{eqnarray}
&& \hspace{-12mm}  \zeta_{A,\vec{0},q} (s) = \zeta_{a,\vec{0},q} (s) +
 \sqrt{\frac{\pi}{a}}\,
\frac{\Gamma (s-1/2)}{\Gamma (s)} \,  \zeta_{\Delta_{p-1},\vec{0},q}
(s-1/2) + \frac{4 \pi^s}{a^{s/2+1/4}\, \Gamma (s)}
\sum_{\vec{n}_2 \in \mbox{\bf Z}^{p-1}}' \nn  \\  &&  \hspace{-8mm}
\sum_{n_1=1}^\infty
 \cos \left( \frac{\pi n_1}{a}  \vec{b}^T  \vec{n}_2 \right)
n_1^{s-1/2}  \left( \vec{n}_2^T \Delta_{p-1}  \vec{n}_2
+q \right)^{1/4 -s/2}
 K_{s-1/2} \left( \frac{2\pi n_1}{\sqrt{a}}
\sqrt{\vec{n}_2^T \Delta_{p-1}  \vec{n}_2 +q} \right). \label{qrpd}
\end{eqnarray}
This is clearly a recurrent formula in $p$, the number of dimensions,
the first term of the recurrence being
\bea
\zeta_{a,\vec{0},q} (s)& =& 2 \sum_{n=1}^\infty \left(
an^2+q\right)^{-s}
= q^{-s} + \sqrt{\frac{\pi}{a}}\, \frac{\Gamma (s-1/2)}{\Gamma (s)} \,
q^{1/2 -s}\nn \\ && + \frac{4\pi^s}{\Gamma (s)} a^{-1/4-s/2} q^{1/4-s/2}
\sum_{n=1}^\infty n^{s-1/2} K_{s-1/2} \left( 2\pi n
\sqrt{\frac{q}{a}} \right). \label{r01}
\eea
To take in these expressions the limit $q\rightarrow 0$ is immediate. One obtains:
 \begin{eqnarray}
&& \hspace{-5mm}  \zeta_{A,\vec{0},0} (s) = 2 a^{-s} \zeta (2s) +
 \sqrt{\frac{\pi}{a}}\,
\frac{\Gamma (s-1/2)}{\Gamma (s)} \,  \zeta_{\Delta_{p-1},\vec{0},0}
(s-1/2) + \frac{4 \pi^s}{a^{s/2+1/4}\, \Gamma (s)}
\sum_{\vec{n}_2 \in \mbox{\bf Z}^{p-1}}' \nn \\  &&  \hspace{-2mm}
\sum_{n_1=1}^\infty
 \cos \left( \frac{\pi n_1}{a}  \vec{b}^T  \vec{n}_2 \right)
n_1^{s-1/2}  \left( \vec{n}_2^T \Delta_{p-1}  \vec{n}_2
 \right)^{1/4 -s/2}
 K_{s-1/2} \left( \frac{2\pi n_1}{\sqrt{a}}
\sqrt{\vec{n}_2^T \Delta_{p-1}  \vec{n}_2 } \right). \label{cspd}
\end{eqnarray}
In the above formulas, $A$ is a $p \times p$ symmetric matrix
$A= \left( a_{ij} \right)_{i,j=1,2, \ldots, p} =A^T$,
$A_{p-1}$  the $(p-1) \times (p-1)$ reduced matrix
$A_{p-1}= \left( a_{ij} \right)_{i,j=2, \ldots, p}$,
$a$ the component $a=a_{11}$, $\vec{b}$ the $p-1$ vector
$\vec{b} =(a_{21}, \ldots, a_{p1})^T = (a_{12}, \ldots, a_{1p})^T$, and
finally, $\Delta_{p-1}$ is the following $(p-1) \times (p-1)$  matrix
 $\Delta_{p-1} =  A_{p-1}- \frac{1}{4a} \vec{b} \otimes \vec{b}$.

Let us now get back to the case $q=0$
starting from the beginning (e.g., the zeta
function given by Eq. (\ref{zf1}), with $q=0$). From that expression,
the recurrence
(\ref{cspd}) can be obtained directly in the same way ---this is
rather obvious, since $q$ plays no role in the derivation. What is not
so obvious
is to realize that the  limit as $q \rightarrow
0$ of Eq. (\ref{qpd1}) is {\it exactly} the recurrent formula
(\ref{cspd}). More precisely, what is obtained in the limit is the
reflected
formula which one gets after using the well known Epstein zeta function
 reflection
\beq
\Gamma (s) Z(s;A) = \frac{\pi^{2s-p/2}}{\sqrt{\det A}}
 \Gamma (p/2-s) Z(p/2-s;A^{-1}),
\eeq
being $Z(s;A)$ the Epstein zeta function \cite{eps1}. After some
thinking,
such result is easy to understand.  Summing up, we have thus checked
that
our formula  (\ref{qpd1}) is valid for {\it any} $q \geq 0$, since it
contains in a hidden way, for $q=0$, the recurrent expression
(\ref{cspd}).

As announced at the beginning, the formulas
derived here  can be considered as generalizations (in more than one
sense) of the Chowla-Selberg (CS) formula. All share the same
 properties that are so much appreciated by number-theoretists as
pertaining to the CS formula. In a way,  these expressions can be viewed
as improved reflection formulas for zeta functions; they are in fact
much better than those
in several aspects. Namely, while a reflection formula connects one
region of the complex plane with a complementary region (with some
intersection) by analytical continuation, the CS formula and our
formulas
are  valid on the {\it whole} complex plane, exhibiting the poles
of the zeta
function and the corresponding residua {\it explicitly}. Even more important,
while a reflection formula is intended to replace the initial
expression of the zeta function ---a power series whose convergence can
be extremely slow--- by another power series with the same type of
convergence, it turns out that the expressions here considered give the
meromorphic extension of the
zeta function, on the whole complex $s$-plane, in terms of an
{\it exponentially decreasing} power series (as was the case with the CS
formula, that one being its most precious property).

Actually, exponential convergence strictly holds
under the condition that
$q\geq 0$. However, the formulas themselves are valid for $q <0$ or even
complex. What is not guaranteed for general $q\in ${\bf C} is the exponential
convergence of
the series, nor its power-like convergence, for that matter. Those analytical
continuations in $q$
must be dealt with specifically, case by case. The physical example
of a field theory with  a chemical
potential falls clearly into this class.
\ms

\ni{\sf 2.3. Particular case $p=2$.}
The above statements apply both for the cases $q >0$ and $q=0$. The
last situation is more involved, however. One is led to employ the
recurrence relation (\ref{cspd}) several
times ($p-1$, in general), what gives rise each time to an additional
series of Bessel functions $K_\nu$    (exponential      convergence).
For completeness, let us write down the corresponding series when $p=2$
explicitly. They are, with $q>0$ \cite{eli2}
 \begin{eqnarray}
&& \hspace{-5mm}  \zeta_E(s;a,b,c;q) = -q^{-s}
+\frac{2\pi q^{1-s}}{(s-1) \sqrt{\Delta}}
 + \frac{4}{\Gamma (s)} \left[
\left( \frac{q}{a} \right)^{1/4}
 \left( \frac{\pi}{\sqrt{qa}} \right)^s
\sum_{n=1}^\infty
n^{s-1/2} K_{s-1/2} \left( 2\pi n \sqrt{\frac{q}{a}} \right) \right.
\nonumber \\ &&  \hspace{6cm} + \sqrt{\frac{q}{a}} \left(2\pi
\sqrt{\frac{a}{q\Delta}} \right)^s
 \sum_{n=1}^\infty n^{s-1} K_{s-1} \left( 4\pi n
\sqrt{\frac{a q}{\Delta}}\right)  \label{cse1}
 \\ && +\left. \sqrt{\frac{2}{a}} (2\pi)^s
 \sum_{n=1}^\infty n^{s-1/2} \cos (\pi n b/a) \sum_{d|n} d^{1-2s} \,
 \left( \Delta + \frac{4aq}{d^2} \right)^{1/4-s/2}
K_{s-1/2}  \left( \frac{\pi n}{a} \sqrt{ \Delta + \frac{4aq}{d^2}}
\right) \right],\nonumber
\end{eqnarray}
where $\Delta = 4ac -b^2>0$, and, with $q=0$, the CS formula \cite{cs}
\begin{eqnarray}
&& \zeta_E(s;a,b,c;0) = 2\zeta (2s)\, a^{-s} + \frac{2^{2s}
\sqrt{\pi}\, a^{s-1}}{\Gamma (s) \Delta^{s-1/2}} \,\Gamma (s
-1/2) \zeta (2s-1)
\nonumber \\ && \hspace{15mm}
+ \frac{2^{s+5/2} \pi^s }{\Gamma (s) \, \Delta^{s/2-1/4}\, \sqrt{a}}
\sum_{n=1}^\infty
 n^{s-1/2}
\sigma_{1-2s} (n) \,
 \cos (\pi n b/a) \,
K_{s-1/2}\left( \frac{\pi n}{a}
\sqrt{ \Delta} \right).
\label{cs1}
\end{eqnarray}
where
$
\sigma_s(n) \equiv \sum_{d|n} d^s,
$
 sum over the $s$-powers of the divisors of $n$.
(There is a missprint in the transcription of  formula  (\ref{cs1}) in
Ref. \cite{dic}). We observe that the rhs's of (\ref{cse1}) and
 (\ref{cs1}) exhibit a
simple pole at $s=1$, with common residue:
\begin{equation}
\mbox{Res}_{s=1}  \zeta_E(s;a,b,c;q) = \frac{2\pi}{\sqrt{\Delta}} =
\mbox{Res}_{s=1}  \zeta_E(s;a,b,c;0).
\end{equation}

\bigskip

\noindent{\large \bf 3. The case of a truncated range}

The most involved case in the family of Epstein-like
 zeta functions corresponds to having to deal with
a {\it truncated} range. This
comes about when one imposes boundary conditions
of the usual Dirichlet or Neumann type \cite{eli2}.
 Jacobi's theta function identity and Poisson's summation
formula are then {\it useless} and no expression in terms of a
convergent series for the analytical continuation to
 values of Re$\, s$ below the abscissa of convergence
can be obtained. The best one gets
 is an {\it asymptotic} series expression. However, the issue of extending
 the CS formula or,  better still, the
most general expression we have obtained before for
 inhomogeneous
 Epstein zeta functions, is not
an easy one. This problem has seldom (if ever)
been properly addressed in the literature.
\ms

\ni{\sf 3.1. Example 1.}
To illustrate the issue, let us consider the following simple example in
one dimension:
\begin{equation}
\zeta_G(s;a,c;q) \equiv \sum_{n=-\infty}^{\infty}
\left[ a(n+c)^2+q
\right]^{-s}, \quad \mbox{Re} \, s >1/2.
\label{g1}
\end{equation}
Associated with this zeta functions, but
 considerably more difficult to treat, is the
truncated series, with indices running from 0 to
$\infty$
\beq
\zeta_{G_t}(s;a,c;q) \equiv \sum_{n=0}^{\infty}
\left[ a(n+c)^2+q
\right]^{-s}, \quad \mbox{Re} \, s >1/2.
\label{g1a}
\end{equation}
 In this case the Jacobi identity is of no use.
 How to proceed then? The only way is to employ
specific techniques of analytic continuation
 of zeta functions \cite{eli2}. The usual method involves three steps
\cite{eli11}. The first step is elementary: to write the
initial series as a Mellin transformed one
\beq
 \sum_{n=0}^{\infty}
\left[ a(n+c)^2+q
\right]^{-s}= \frac{1}{\Gamma (s)} \sum_{n=0}^{\infty} \int_0^\infty
dt \, t^{s-1} \exp\left\{ -[ a(n+c)^2+q]t \right\}.
\label{g1b}
\end{equation}
The second is to expand in power series part of the exponential, while
leaving always a converging exponential factor,
 \beq
 \sum_{n=0}^{\infty} \left[ a(n+c)^2+q \right]^{-s}= 
\frac{1}{\Gamma (s)} \sum_{n=0}^{\infty} \int_0^\infty
dt \,  \sum_{m=0}^{\infty} \frac{(-a)^m}{m!} (n+c)^{2m} t^{s+m-1}
e^{-qt}. \label{g1c}
\end{equation}
The third and most difficult step is to interchange the order of the
two summations ---with the aim to obtain a series of zeta functions---
what means transforming the second series into
an integral along a path on the complex plane, that has to be closed
into a circuit (the sum over poles inside reproduces the original series), 
with a part of it being sent to
infinity. Usually, after interchanging the first
 series and the integral, there is a contribution of this part
of the circuit at infinity, what provides in the end an {\it additional}
contribution
to the trivial commutation. More important, what one obtains in general
through this process is
{\it not} a convergent series of zeta functions, but an asymptotic
series \cite{eli2}. That is, in our example,
\beq
\sum_{n=0}^{\infty} \left[ a(n+c)^2+q \right]^{-s}\sim
\sum_{m=0}^{\infty} \frac{(-a)^m\Gamma (m+s)}{m!\, \Gamma (s) \,
q^{m+s}} \zeta_H (-2m, c) + \, \mbox{additional terms}.
\eeq
Being more precise, as outcome of the whole process 
 we obtain the following result for the analytic
continuation of the zeta function \cite{elif1}:
\begin{eqnarray}
&& \hspace{-4mm} \zeta_{G_t}(s;a,c;q)
 \sim \left(\frac{1}{2} -c \right) q^{-s} + \frac{q^{-s}}{\Gamma (s)}
\sum_{m=1}^{\infty}
\frac{(-1)^m \Gamma (m+s)}{m!} \left( \frac{q}{a} \right)^{-m}
\zeta_H (-2m, c)  \label{if11} \\ &&   +
\sqrt{\frac{\pi}{a}} \, \frac{\Gamma (s-1/2)}{2\Gamma (s)} q^{1/2 -s}
+\frac{2\pi^s}{\Gamma (s)} a^{-1/4-s/2} q^{1/4-s/2}
 \sum_{n=1}^\infty
n^{s-1/2} \cos (2\pi nc) K_{s-1/2} (2\pi n\sqrt{q/a}).  \nonumber
\end{eqnarray}
(Note that this expression reduces to Eq. (\ref{r01}) in the limit
$c \rightarrow 0$.)
The first series on the rhs is asymptotic \cite{eli11,8}.
Observe, on the other hand, the singularity structure of this zeta
function. Apart from the pole at $s=1/2$,  there is a whole sequence of
poles at the negative real axis, for $s= -1/2, -3/2, -5/2, \ldots$,
with residua:
\beq
\mbox{Res}_{s=1/2-j} \zeta_{G_t}(s;a,c;q) = \frac{(2j-1)!!\, q^j}{
j!\, 2^j \sqrt{a}}, \ \ \ j=0,1,2, \ldots
\eeq
\ms

\ni{\sf 3.2. Example 2.}
As a second example, in order to obtain the analytic continuation to
  Re $s  \leq 1$  of the
 truncated inhomogeneous Epstein zeta
function in two dimensions,
\begin{equation}
 \zeta_{E_t}(s;a,b,c;q) \equiv \sum_{m,n =0}^\infty
(am^2+bmn+cn^2+q)^{-s},
\end{equation}
we can  proceed in two ways: either by direct
calculation following the three steps as explained above
or else by using the final formula for the  Epstein zeta
function in  one dimension (example 1)
 recurrently. In both cases the end result  is the same:
 \begin{eqnarray}
&& \hspace{-5mm}   \zeta_{E_t}(s;a,b,c;q)
 \equiv \sum_{m,n =0}^\infty
(am^2+bmn+cn^2+q)^{-s}   \nonumber \\
&&  \hspace{-2mm}  \sim \, \frac{(4a)^s}{\Gamma (s)} \sum_{m,n =1}^\infty
\frac{(-1)^m \Gamma (m+s)}{m!} (2a)^{2m} (\Delta n^2 +4aq)^{-m-s}
\zeta_H \left(-2m; \frac{bn}{2a} \right)  \nonumber \\
&& -\frac{b\, q^{1-s}}{(s-1)\Delta \Gamma (s-1)}
 \sum_{n =0}^\infty \frac{(-1)^n \Gamma (n+s-1)B_n}{n!}
 \left( \frac{4aq}{\Delta} \right)^{-n}
  + \frac{q^{-s}}{4}
 +\frac{\pi q^{1-s}}{2(s-1) \sqrt{\Delta}} \nn \\ &&
\hspace{-8mm} + \frac{1}{4}
\left( \sqrt{\frac{\pi}{a}} + \sqrt{\frac{\pi}{c}} \right)
\frac{\Gamma (s-1/2)}{\Gamma (s)} q^{1/2-s}
 + \frac{1}{\Gamma (s)} \left[ 2
\left( \frac{q}{a} \right)^{1/4}
 \left( \frac{\pi}{\sqrt{qa}} \right)^s
\sum_{n=1}^\infty
n^{s-1/2} K_{s-1/2} \left( 2\pi n \sqrt{\frac{q}{a}} \right)
\right. \nn
 \\ &&\hspace{-8mm}  +\left( \frac{aq}{\Delta} \right)^{1/4}
 \left( \pi \sqrt{\frac{a}{q\Delta}} \right)^s
\sum_{n=1}^\infty
n^{s-1/2} K_{s-1/2} \left( 2\pi n \sqrt{\frac{aq}{\Delta}} \right)
+ \sqrt{\frac{q}{a}} \left(2\pi
\sqrt{\frac{a}{q\Delta}} \right)^s
 \sum_{n=1}^\infty n^{s-1} K_{s-1} \left( 4\pi n
\sqrt{\frac{a q}{\Delta}}\right)  \nonumber
 \\ &&  \hspace{-8mm} + \left. \sqrt{\frac{2}{a}} (2\pi)^s
 \sum_{n=1}^\infty n^{s-1/2} \cos (\pi n b/a) \sum_{d|n} d^{1-2s} \,
 \left( \Delta + \frac{4aq}{d^2} \right)^{1/4-s/2}
K_{s-1/2}  \left( \frac{\pi n}{a} \sqrt{ \Delta + \frac{4aq}{d^2}}
\right) \right].\label{cser1}
\end{eqnarray}
 The first series on the rhs is in general
asymptotic, although it converges for a  wide range of values of
the parameters. The second series is always asymptotic and
its first term contributes to the pole at $s=1$. As in the case of
Eq. (\ref{cse1}), the pole structure is here
explicit, although  much more elaborate.
Apart from the pole at $s=1$, whose residue is
\beq
\mbox{Res}_{s=1} \zeta_{E_t}(s;a,b,c;q) = \frac{\pi}{2\sqrt{\Delta}} -
\frac{b}{\Delta},
\eeq
 there is here also a sequence
of poles at $s=\pm 1/2, -3/2, -5/2, \ldots$, with residua:
\beq
\mbox{Res}_{s=1/2-j} \zeta_{E_t}(s;a,b,c;q) = \frac{(2j-1)!!\, q^j}{
j!\, 2^{j+2}} \left( \frac{1}{\sqrt{a}} + \frac{1}{\sqrt{c}}\right), \
\ \ j=0,1,2, \ldots
\eeq

The formula above, Eq. (\ref{cser1}), is really imposing and hints
already towards the conclusion that
the derivation of a general expression in $p$ dimensions for the zeta
function considered in Sect. 2 but with a truncated range is not an easy
task.

\bigskip

\noindent{\large \bf 4. Some uses of the formulas}

These formulas are very powerful expressions in order to determine the
analytic
structure of generalized inhomogeneous Epstein type zeta functions, to
obtain specific values of these zeta functions at different points,
and from there, in particular, the Casimir effect and heat kernel
coefficients,
and also in order to calculate derivatives of the zeta function, and
from
them, in particular, the associated determinant. Notice that obtaining
derivatives of the formulas in Sect. 2 presents no problem. Only for
truncated zeta functions (Sect. 3) the usual care must be taken when dealing with
asymptotical expansions.
We shall illustrate these uses with three specific applications.
\ms

\ni{\sf 4.1. Application 1.}
In a recent paper by R. Bousso and S. Hawking \cite{bh1}, where the
{\it trace
anomaly} of a dilaton coupled scalar in two dimensions is calculated, the
zeta function method is employed for obtaining the one-loop effective
action, $W$, which is given by the well known expression
\beq
W= \frac{1}{2} \left[\zeta_A (0) \ln \mu^2 + {\zeta_A}'(0) \right],
\eeq
with $\zeta_A (s) =$ tr $A^{-s}$. In conformal field theory and in a
Euclidean background manifold of toroidal topology, the eigenvalues of
$A$ are found perturbatively (see \cite{bh1}), what leads one to
consider the following zeta function:
\beq
\zeta_A (s) = \sum_{k,l=-\infty}^\infty (\Lambda_{kl})^{-s},
\eeq
with the eigenvalues $\Lambda_{kl}$ being given by
\beq
\Lambda_{kl} = k^2+l^2+\frac{\epsilon^2}{2} +
\frac{\epsilon^2}{2(4l^2-1)},
\eeq
where $\epsilon$ is a perturbation parameter. It can be shown that
the
integral of the trace anomaly is given by the value of the zeta function
at $s=0$. One barely needs
to follow the several pages long discussion in \cite{bh1}, leading to
the calculation of this
value, in order to appreciate the power of the formulae of the
preceding section. In fact, to begin with, no mass term needs to be
introduced to arrive at the result and no limit mass $\rightarrow 0$
needs
to be taken later. Using  binomial expansion (the same as in Ref.
\cite{bh1}), one gets \beq
\zeta (s) = \sum_{k,l=-\infty}^\infty \left(k^2+l^2+\frac{\epsilon^2}{2}
\right)^{-s} - \frac{\epsilon^2 s}{2}  \sum_{k,l=-\infty}^\infty
 \left(k^2+l^2+\frac{\epsilon^2}{2} \right)^{-1-s} (4l^2-1)^{-1}.
\eeq
From Eq. (\ref{cse1}) above, the first zeta function gives, at $s=0$,
{\it exactly}: $-\pi \epsilon^2/2$. And this is the whole result (which
does coincide with the one obtained in \cite{bh1}),
since the second term has no pole at $s=0$ and provides no contribution.
\ms

\ni{\sf 4.2. Application 2.}
Another direct application is the calculation of the {\it Casimir energy
density} corresponding to a massive scalar field on a general, $p$
dimensional toroidal manifold (see \cite{ke1}).
In the spacetime ${\cal M} =$ {\bf R}
$\times \Sigma$, with $\Sigma = [0,1]^p/$$\sim$, which is
topologically equivalent
to the $p$ torus, the Casimir energy density for a massive scalar field
is given directly by Eq. (\ref{qpd1}) at $s=-1/2$, with $q=m^2$ (mass of
the
field), $\vec{b} =\vec{0}$, and $A$ being the matrix of the metric $g$ on
$\Sigma$, the general $p$-torus:
\beq
E^C_{{\cal M}, m} = \zeta_{g, \vec{0}, m^2} (s=-1/2).
\eeq
The components of $g$ are, in fact, the
coefficients of
the different terms of the Laplacian, which is the relevant operator in
the Klein-Gordon field equation. The massless case is also obtained,
with the same specifications, from the corresponding formula  Eq.
(\ref{cspd}). In both cases no extra calculation needs to be done,
and the physical results follows from a mere
{\it identification} of the components of the matrix $A$ with those of the
metric tensor of the manifold in question \cite{ke1}. Very much related with
this application but more involved and ambitious  is the calculation of
vacuum  energy densities corresponding to spherical configurations and
the bag model (see
\cite{bekl1,wipf12}, and the many references therein).
\ms

\ni{\sf 4.3. Application 3.}
A third application consists in calculating the {\it determinant} of a
differential operator, say the Laplacian on a general $p$-dimensional
torus. A very important problem related with this issue is that
 of the associated anomaly (called the multiplicative or noncommutative 
anomaly) \cite{a1}.
 To this end the derivative of the zeta function at $s=0$ has to
be obtained. From Eq. (\ref{qpd1}), we get
\bea
\hspace{-15mm} {\zeta'}_{A,\vec{c},q} (0) &=&
 \frac{4 (2q)^{p/4}}{\sqrt{\det A}}
\sum_{\vec{m} \in \mbox{\bf Z}^p_{1/2}}' \frac{\cos (2\pi
 \vec{m}\cdot \vec{c})}{ \left( \vec{m}^T A^{-1} \vec{m}
\right)^{p/4}} \, K_{p/2} \left( 2\pi \sqrt{2q \,
 \vec{m}^T A^{-1} \vec{m}}\right) \nn \\ && \hspace{20mm} + \left\{
\brr{ll} \dsp\frac{(2\pi )^{p/2} \Gamma (-p/2) q^{p/2}}{
\sqrt{\det A}}, & p \ \mbox{odd}, \\
\dsp\frac{(-1)^k(2\pi )^k q^k}{k!\,
\sqrt{\det A}} \, \left[ \Psi (k+1) +\gamma -\ln q \right], &
p=2k \ \mbox{even}, \err \right.
 \label{qpdd1}
\eea
and, from here, det $A$ = exp $-\zeta_A'(0)$. For $p=2$, we have
explicitly:
\bea
\det A(a,b,c;q) &=& e^{2\pi (q- \ln q)/\sqrt{\Delta}} \left( 1-
e^{-2\pi \sqrt{q/a}} \right) \exp \left\{ -4 \sum_{n=1}^\infty
\frac{1}{n} \left[ \sqrt{\frac{a}{q}} \ K_1 \left( 4\pi n \sqrt{
\frac{aq}{\Delta}} \right) \right. \right. \nn \\ && +
\left. \left. \cos (\pi n b /a) \sum_{d|n} d \, \exp \left(-
\frac{\pi n}{a} \sqrt{ \Delta + \frac{4aq}{d^2}} \right) \right]
\right\}. \eea
In the homogeneous case (CS formula) we obtain for the determinant:
\beq
\det A(a,b,c) = \frac{1}{a} \exp \left[ -4 \zeta'(0) -
\frac{\pi \sqrt{\Delta}}{6a} -4 \sum_{n=1}^\infty \frac{\sigma_1(n)}{n}
\cos (\pi n b /a) e^{-\pi n \sqrt{\Delta} /a} \right],
\eeq
or, in terms of the Teichm\"uller coefficients, $\tau_1$ and $\tau_2$,
of the metric tensor (for
the metric, $A$, corresponding to the general torus in two dimensions):
\beq
\det A(\tau_1, \tau_2) = \frac{\tau_2}{4\pi^2|\tau|^2} \exp \left[ -4
\zeta'(0) -
\frac{\pi \tau_2}{3|\tau|^2} -4 \sum_{n=1}^\infty \frac{\sigma_1(n)}{n}
\cos \left(\frac{2\pi n \tau_1}{|\tau|^2} \right) e^{-\pi n
\tau_2/|\tau|^2} \right].
 \eeq
Needless to mention, all the good properties of the expression for the
zeta function are just transferred to  the associated
determinants, which
are thus given, on its turn, in terms of very quickly convergent
series.

\vspace{3mm}

\noindent{\bf Acknowledgments}.
The author is indebted with Andreas Wipf, Michael Bordag, Klaus
Kirsten and Sergio Zerbini for enlightening
discussions and with the members of the Institutes
of Theoretical Physics of the Universities of Jena and Leipzig,
where the main part of this work was done, for
very kind hospitality.
 This investigation has been  supported by
CIRIT (Generalitat de Catalunya), DGICYT (Spain) and by the
German-Spanish program Acciones Integradas, project HA1996-0069.

\end{document}